\documentclass[twoside]{photon2007}
\usepackage[latin1]{inputenc}
\usepackage[dvips]{graphicx,epsfig,color}
\usepackage{wrapfig,rotating}
\usepackage{amssymb,amsmath,array}

\begin{document}
\title{Transversity Measurements at COMPASS} 
\author{C. Schill on behalf of the COMPASS Collaboration
\vspace{.3cm}\\
Physikalisches Institut der Universit\"at Freiburg \\
Hermann-Herder Str. 3, D-79104 Freiburg, Germany.
\vspace{.1cm}\\
}

\maketitle

\begin{abstract}  The measurement of transverse spin effects in semi-inclusive
deep-inelastic scattering is an important part of the COMPASS physics program. 
From the analysis of the 2002-2004 data, new results for the transverse target
spin asymmetry of z-ordered identified pion and kaon pairs  are presented
\cite{url}.  In addition, a first result for the transverse target spin
asymmetry of exclusively produced $\rho^0$ mesons on the deuteron is shown. 
\end{abstract}

\section{Introduction} Single spin asymmetries in semi-inclusive deep-inelastic
scattering off transversely polarized nucleon targets have been under intense
experimental and theoretical investigation over the past few years
\cite{COMPASS, HERMES, Bacchetta}. They provide new insights into QCD and the
nucleon structure. For instance, they allow the determination of the third
yet-unknown leading-twist quark distribution function $\Delta _Tq(x)$, the
transversity distribution \cite{Collins, Artru}. Additionally, they give
insight into the parton transverse momentum distribution and angular momentum
\cite{Jaffe}.

COMPASS exploited three different "quark polarimeters" to access transversity:
For hadron production in deep-inelastic scattering the azimuthal asymmetry of
the produced hadrons was measured. This asymmetry gives access to transversity
by the Collins mechanism. Another probe to access transversity is the
interference fragmentation function in the production of hadron pairs
\cite{Bianconi, Radici}. Finally, for the production of $\Lambda$
baryons the transverse spin of quarks is transferred to the transverse lambda
polarization, which is determined experimentally.

In hard exclusive production of $\rho^0$ mesons on a transversely polarized
target, a different aspect of the spin structure of the nucleon can be probed.
Since few years it is well established that hard meson production is a very
good candidate to study the universal generalized parton distributions (GPDs)
\cite{Vanderhaegen, Goeke, Diehl}. These distributions describe in the most
complete way the nucleon structure and encode fundamental information in
particular about the angular momentum carried by partons and about their
spatial distribution. 

At the COMPASS experiment at CERN  all channels mentioned above have been
studied. The following part will focus on two new results, the Collins
asymmetry  for identified z-ordered hadron pairs and the target spin asymmetry
for exclusively produced  $\rho^0$ mesons.

\section{Two-hadron asymmetry} 
At leading twist, the fragmentation function ($FF$) of a polarized quark into a pair
of hadrons is expected to be of the form 
\begin{equation*}
FF=D_q^{2h}(z, M_h^2) +  H_1^\sphericalangle(z,M_h^2) \sin\theta \sin \phi_{RS},
\end{equation*}
where $M_h$ is the invariant mass of the hadron pair and $z=z_1+z_2$ is the
fraction of available energy carried by the two hadrons. $D_q^{2h}(z, M_h^2)$ is
the unpolarized fragmentation function into two hadrons. The angles $\theta$
 and $\phi_{RS}$ are defined according to Ref. \cite{Artru2}. 
$\phi_{RS}=\phi_R+\phi_{S}-\pi$ is the sum  of the azimuthal angle $\phi_R$ of
a plane containing the two hadrons and the  azimuthal angle $\phi_S$ of target
spin vector with respect to the lepton scattering plane. $\theta$ is the
polar angle of the first hadron in the two-hadron center-of-mass frame with
respect to the direction of the summed hadron momentum
$\overrightarrow{p}_h=\overrightarrow{p}_1+\overrightarrow{p}_2$. 
Within the COMPASS acceptance, $\theta$ peaks close to $\pi/2$ with
$<\sin\theta>\approx 0.95$. The following results are obtained by integrating over $\sin\theta$. 
The number of hadron pairs in a bin of  $x$, 
$z$ or $M_h$ is given by 
\begin{equation*}
N^\pm(\phi_{RS})=N_0\cdot(1\pm A_{UT}^{\sin\phi_{RS}}\cdot \sin \phi_{RS}),
\end{equation*}
where $\pm$ refers to the transverse target spin orientation and $N_0$ is the
mean number of detected hadron pairs averaged over $\sin\phi_{RS}$. 
From the angular distribution of the hadron pairs, one can thus measure the
asymmetry 
\begin{equation*}
A_{RS}=\frac{1}{f P_T D}\cdot A_{UT}^{\sin\phi_{RS}},
\end{equation*}
where $f\approx 0.38$ is the target dilution factor, $P_T$ 
the target polarization and $D$ the depolarization factor given by
 $D=(1-y)/(1-y+y^2/2)$, where  $y$ is the fractional energy
transfer of the lepton.

The measured asymmetry can be factorized into a convolution of the transversity
distribution $\Delta_Tq(x)$ of the quarks of flavor $q$ and the interference
fragmentation  $H_1^\sphericalangle(z,M_h^2)$ \cite{Artru2}:
\begin{equation*}
A_{RS}=\frac{\Sigma_q \;e_q^2 \cdot \Delta_Tq(x) \cdot H_1^\sphericalangle(z,M_h^2)}
{\Sigma_q \;e_q^2 \cdot q (x) \cdot D_q^{2h}(z, M_h^2)},
\end{equation*}
summed over all quark flavors $q$.

\section{Event selection} 
\label{selection}
The data discussed here have been taken in the years
2003 to 2004 at the COMPASS experiment at CERN. It scatters a $160~GeV$ $\mu^+$ beam
on a transversely polarized solid state $^6$LiD target. The scattered muons and
the produced hadrons are detected in a $50$~m long large-acceptance forward
spectrometer with excellent particle identification capabilities. A large scale
Ring Imaging Cherenkov Detector (RICH) is used to distinguish pions, kaons and
protons \cite{Experiment}.

 The event selection was done in the same way as in the previous analysis of
the Collins and Sivers asymmetries for single hadrons \cite{COMPASS}. For the
selection of the DIS event sample, kinematic cuts of the squared four-momentum
transfer $Q^2> 1$~(GeV/c)$^2$, the hadronic invariant mass $W> 5$~GeV/c$^2$ and
the fractional energy transfer of the muon $0.1 <y <0.9$ were applied.

Hadron pairs originating from the primary vertex are selected. The hadrons are
separated into $\pi$ and $K$ pairs. A selection cut of $z_{1/2}>0.1$ suppresses
hadrons from the target fragmentation. The two leading hadrons have been
selected ordered according to their fractional energy, taking the first hadron as
the most energetic one. A cut on the sum  $z=z_1+z_2<0.9$  rejects exclusively
produced $\rho$ and $\phi$ mesons. 

By combining data from both target cells as well as from sub-periods with opposite
target polarization in a double ratio product described in detail in Ref.
\cite{COMPASS}, the acceptance function of the spectrometer cancels out and the
azimuthal asymmetry $A_{RS}(x, z, M_h) $ is  extracted by a fit to the data. In various
studies, it was shown that  systematic effects of the measurement are considerably
smaller than the statistical uncertainty of the data.

\section{Results} 
\begin{figure}
\includegraphics[clip,bb=0 5 567 322,height=0.18\textheight, width=\columnwidth]{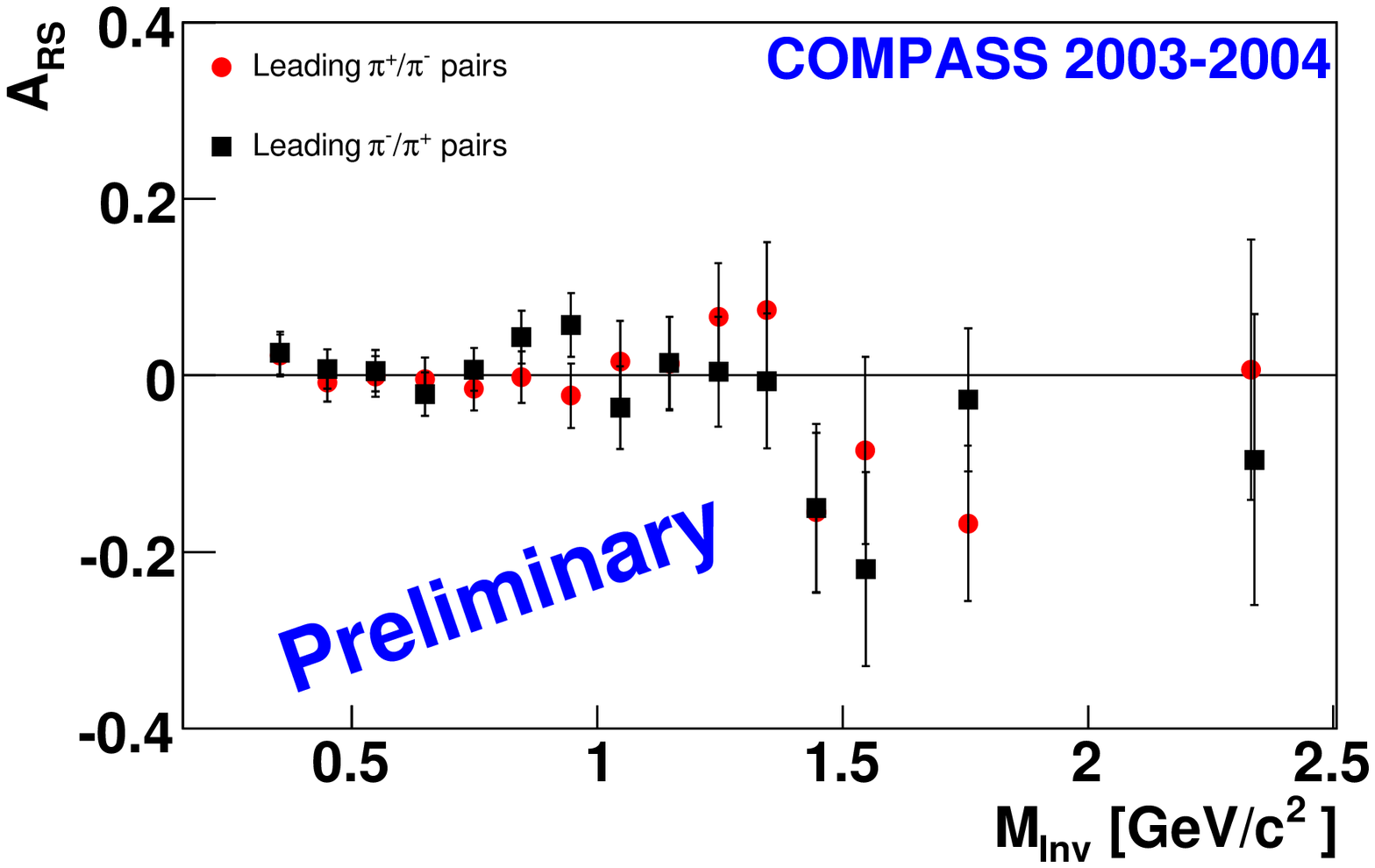}\\[-0cm]
\includegraphics[clip,bb=0 5 567 322,height=0.18\textheight, width=\columnwidth]{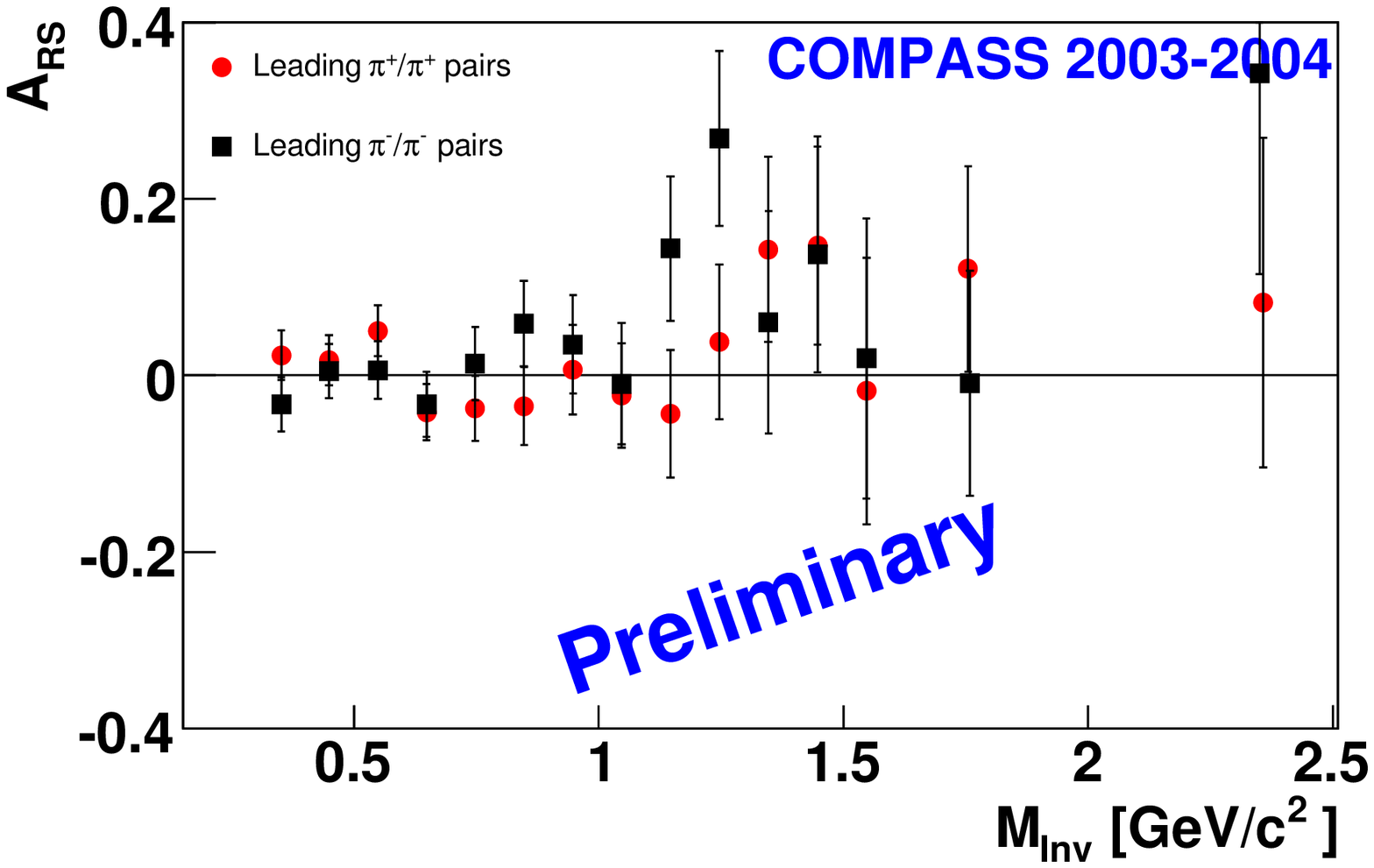}\\[-0cm]
\includegraphics[clip,bb=0 5 567 322,height=0.18\textheight, width=\columnwidth]{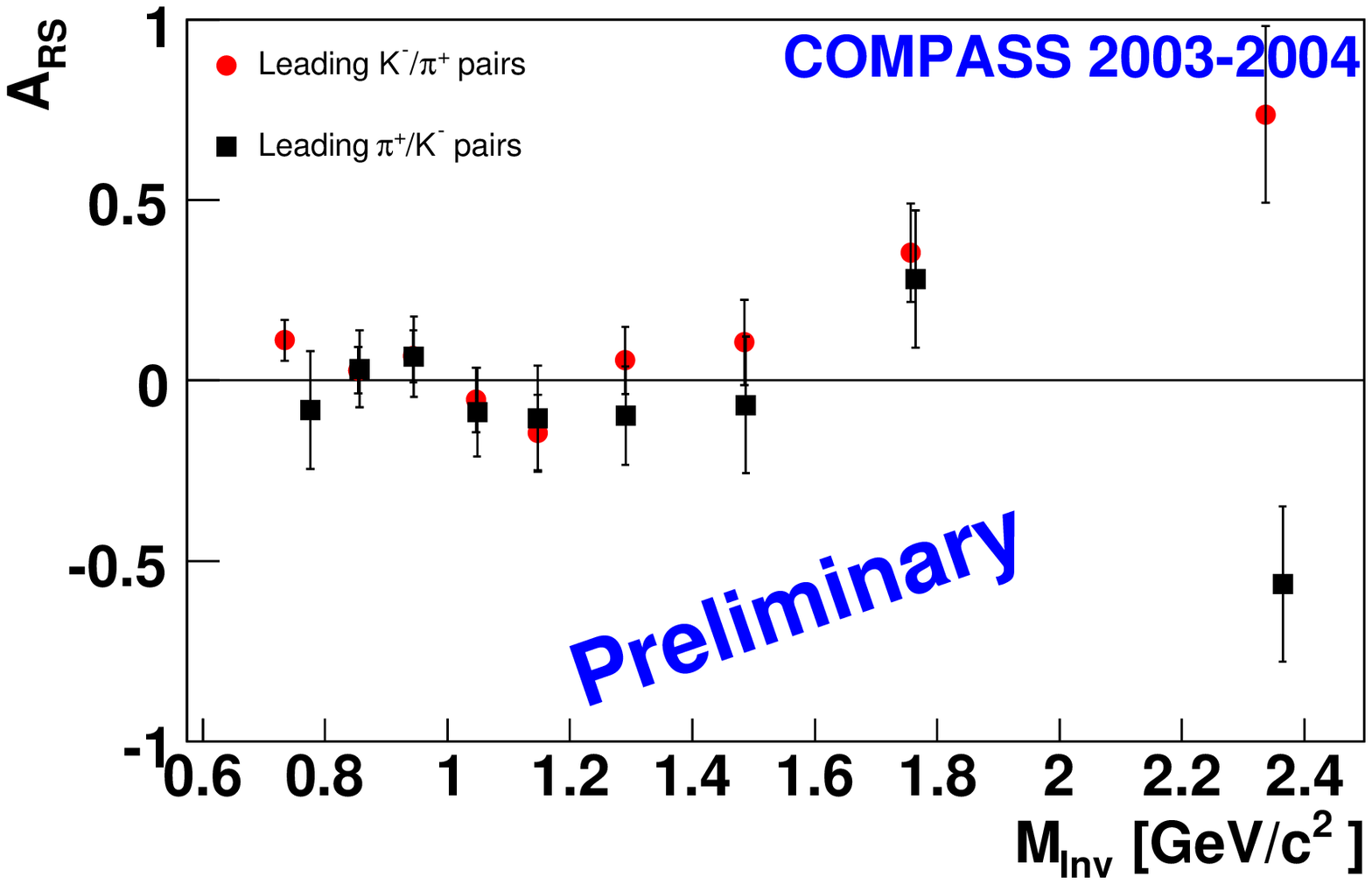}\\[-0cm]
\includegraphics[clip,bb=0 5 567 322,height=0.18\textheight, width=\columnwidth]{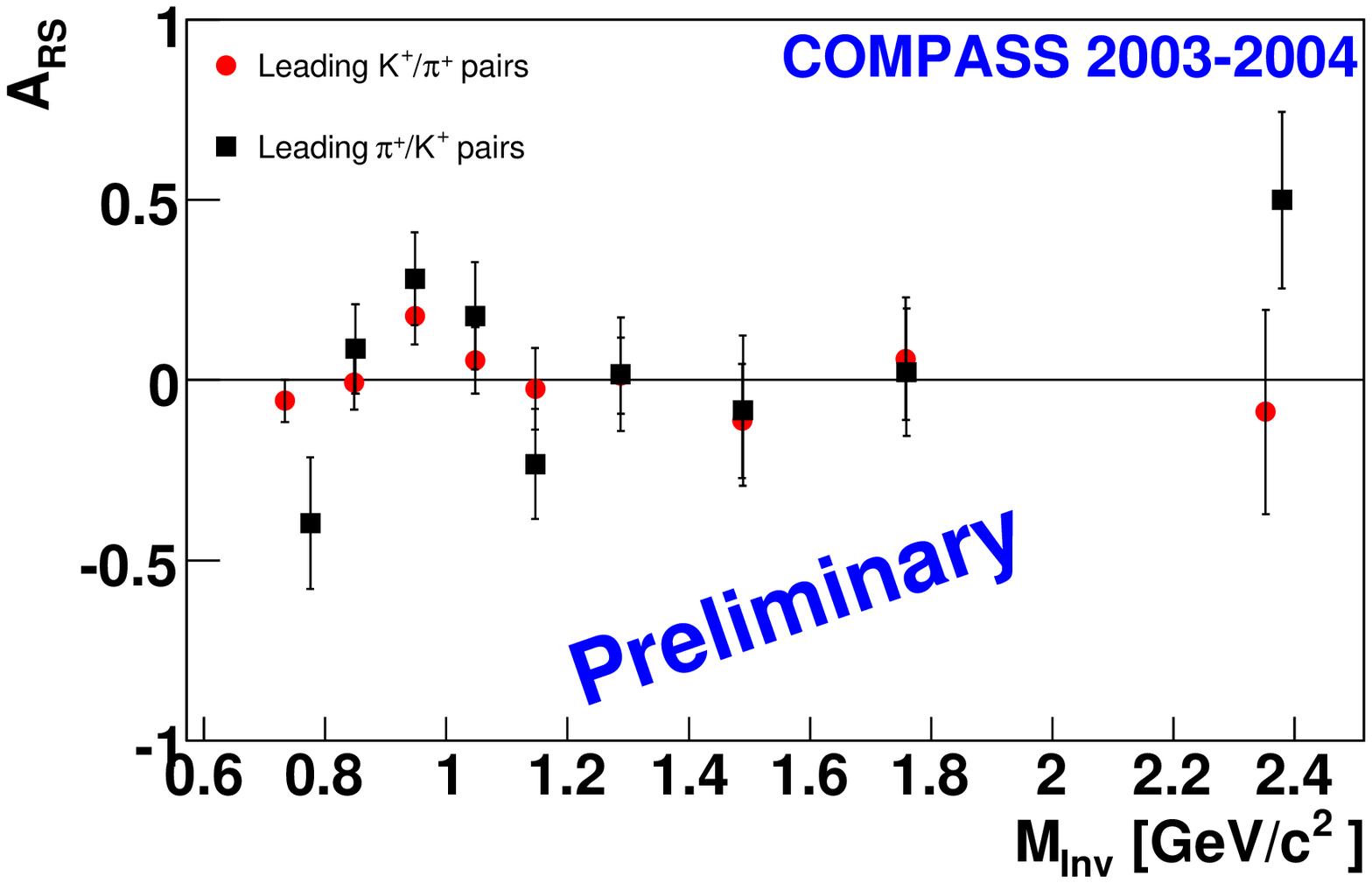}\\[-0cm]
\includegraphics[clip,bb=0 5 567 322,height=0.18\textheight, width=\columnwidth]{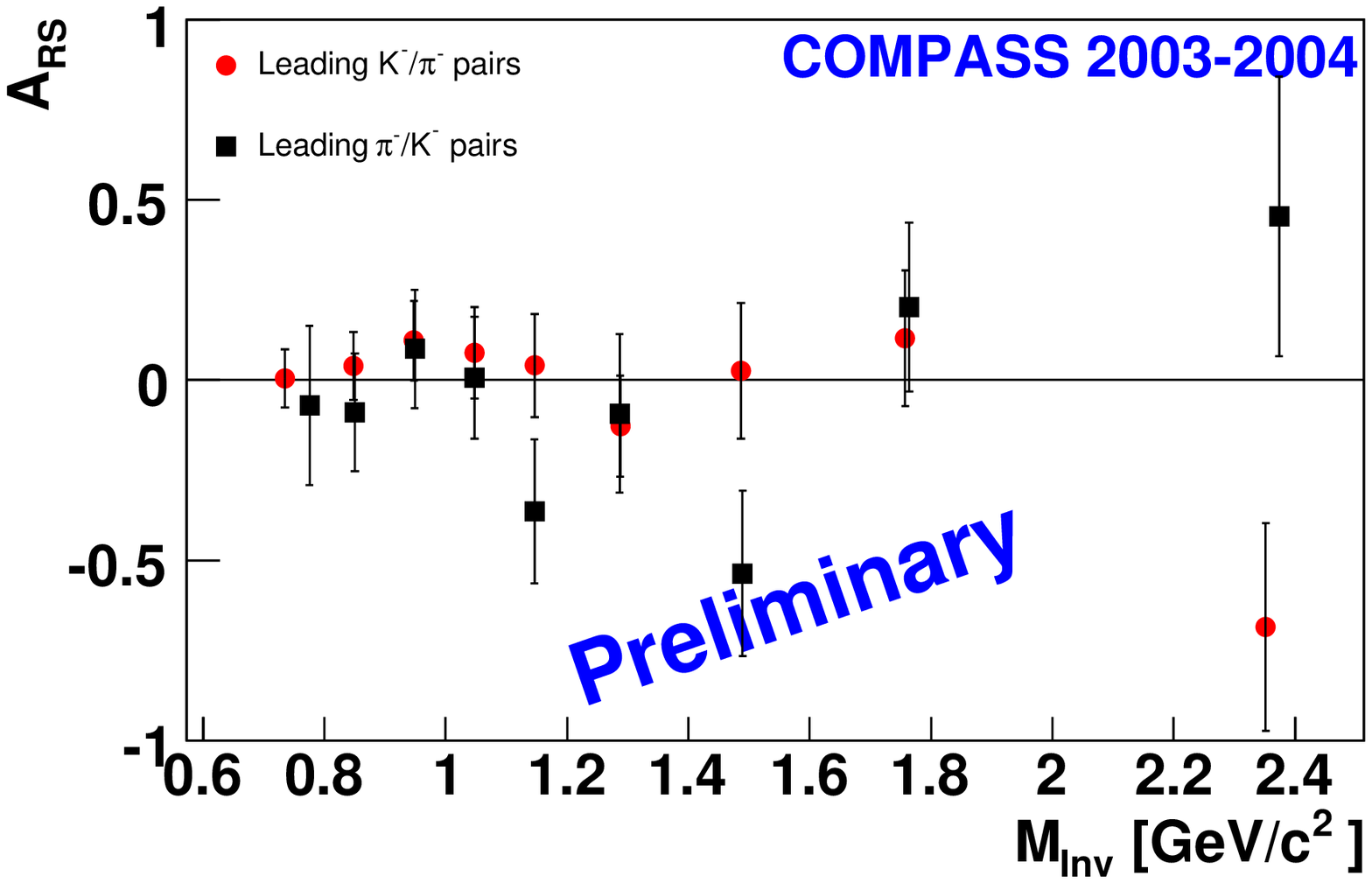}\\[-0.7cm]
\caption{Asymmetries $A_{RS}$ for identified leading $\pi\pi$ pairs with
opposite 
 ({\it top}) and equal ({\it second}) charge, for identified $K \pi$ pairs 
with opposite ({\it third}) and equal pos. ({\it fourth}) and neg. charge ({\it bottom panel}) versus $M_{inv}$.}
\label{Result1}
\end{figure}

In figure \ref{Result1} the results for the target single spin
asymmetry $A_{RS}$ for identified leading z-ordered $\pi\pi$ and $K \pi$ pairs are shown. The
asymmetries are plotted as a function of the hadron pair invariant mass $M_{inv}$.
The measured asymmetries are small and compatible with zero within the
statistical precision of the data points. They do not show a significant
dependence on the  kinematic variables $M_{inv}$, or $x$ and $z$, respectively.

\section{Discussion}
In several theoretical models, predictions have been made for the measured
asymmetries $A_{RS}(x, z, M_h)$ for pions or unidentified hadrons on a
deuteron target \cite{Bacchetta, Bianconi}. The expected values of the
asymmetry are generally small and below $1\%$. The small signal is attributed
in these calculations to a partial cancellation of the asymmetries originating
from scattering on $up$ and $down$ quarks of the proton and neutron in the
isoscalar deuteron target. In 2007, COMPASS is taking data with a transversely
polarized proton target, where the asymmetries are expected to be larger
\cite{HERMES, Bacchetta}.
Together with the deuteron data presented here, a separation of the asymmetries
originating from $up$ and $down$ quarks shall then become possible.

\section{Exclusive $\rho^0$ production} Hard meson production has been shown to
be one possible way to access generalized parton distributions. Here we have
studied the exclusive production of $\rho^0$ mesons on a transversely polarized
deuteron target \cite{Jasmin}. 

The analysis is based on the complete COMPASS transverse data set from
2002-2004. For the selection of DIS events, the same cuts as described in
section \ref{selection} have been applied. To select exclusive $\rho^0$, a
missing energy cut of $-2.5$~GeV $< E_{miss} < 2.5$~GeV has been applied.
$\rho^0$ mesons have been selected in a invariant mass window of $300$~MeV around
the $\rho^0$ mass. A cut on the squared transverse momentum $p_t^2$ of the
$\rho^0$ meson
in the range between $0.01$ and $0.5$~GeV/c$^2$ has been applied, where the
lower limit ensures an accurate measurement of the azimuthal angle and the upper
one suppresses non-exclusive background. In total $270k$  $\rho^0$ mesons have
been reconstructed. The selected data sample contains $\rho^0$ produced
incoherently and coherently on the deuteron target as well.

\begin{figure*}[t]
\includegraphics[bb=25 0 540 407,clip, width=0.329\textwidth]{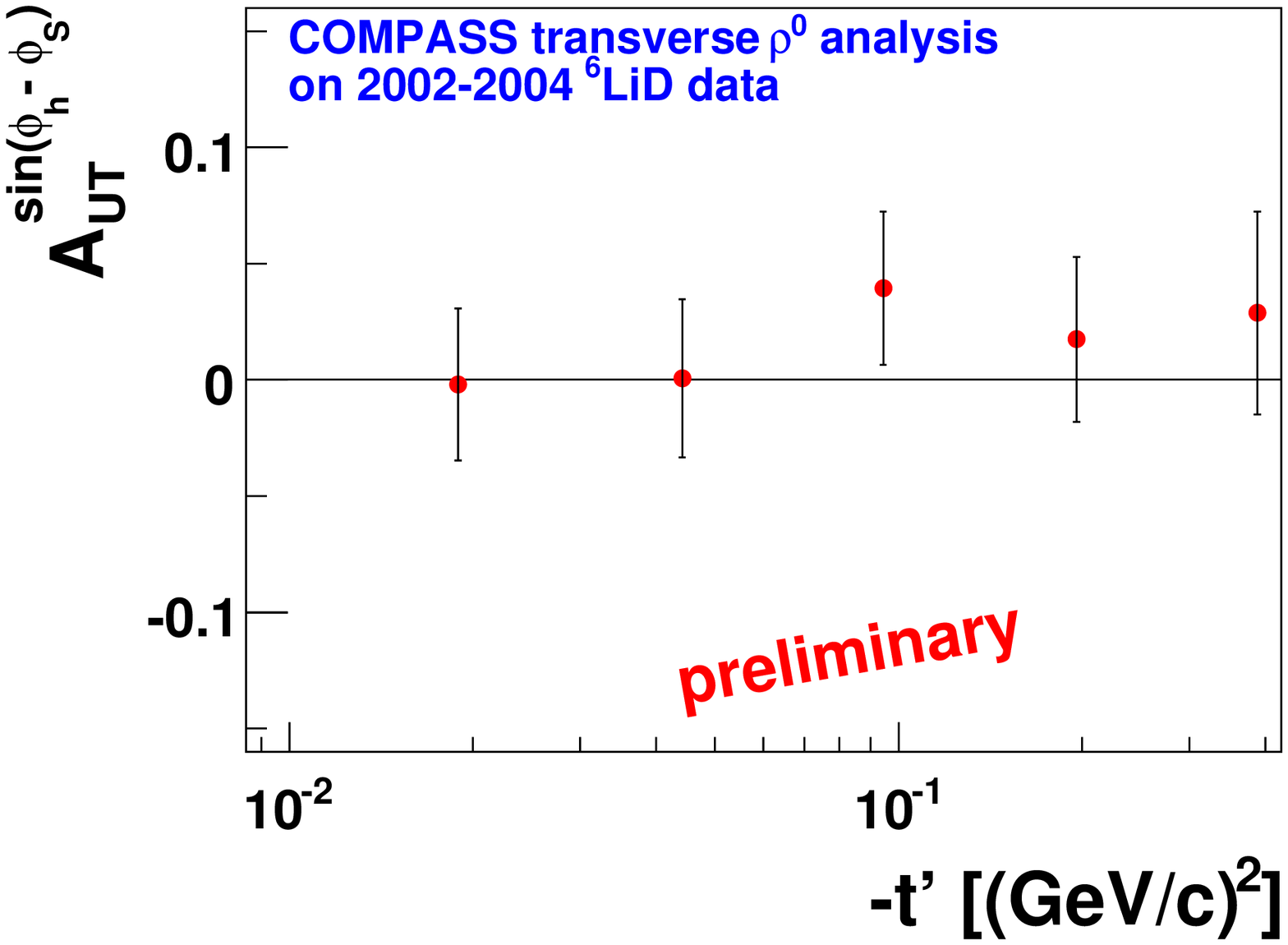}
\includegraphics[bb=25 0 540 407,clip, width=0.329\textwidth]{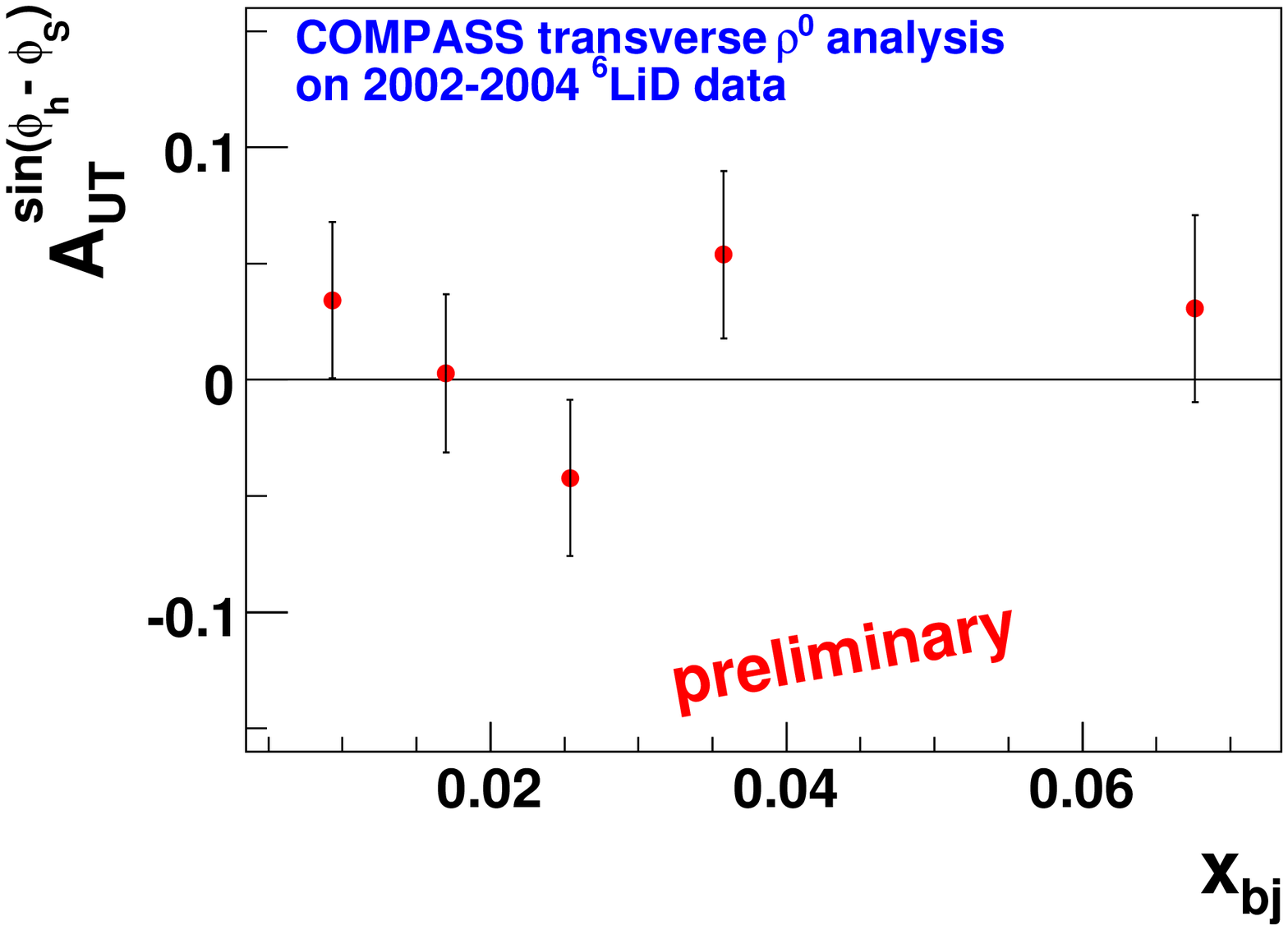}
\includegraphics[bb=25 0 540 407,clip,width=0.329\textwidth]{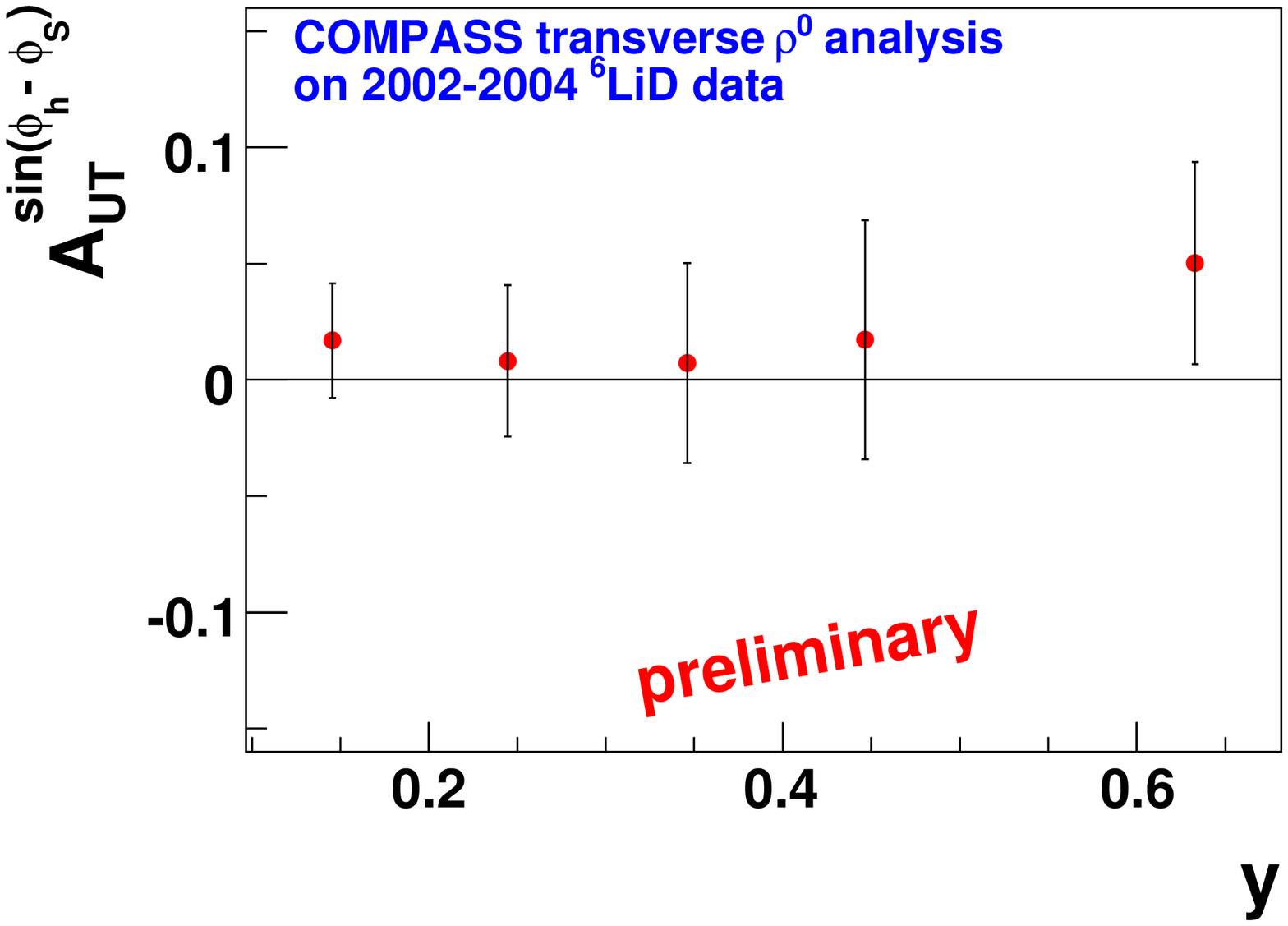}\vspace*{-0.4cm}
\caption{Target single spin asymmetry $A_{UT}$ for exclusive $\rho^0$ 
production on the deuteron.}
\label{Result2}
\end{figure*}

An azimuthal asymmetry $A^{\sin\eta}$ has been determined with
respect to the angle $\eta=\phi_\rho-\phi_S$, where $\phi_\rho$ is the azimuthal
angle of the $\rho^0$ meson with respect to the lepton scattering plane.
 The transverse target spin asymmetry
has been calculated from the measured asymmetry $A^{\sin\eta}$
\begin{equation*}
A_{UT}^{\sin\eta}=\frac{1}{f\cdot P_T}A^{\sin\eta},
\end{equation*}
where $f=0.38$ is the dilution factor and $P_T$ the mean target polarization.

Results for the asymmetry $A_{UT}$ as a function of the transferred four momentum
above threshold $t'$, of $x$ and of the fractional photon energy transfer 
$y$ are shown in figure \ref{Result2}.

\section{Discussion of $\rho^0$ results}

For longitudinal virtual photons, factorization of the exclusive vector meson
production cross section into a hard part sensitive to GPDs and a fragmentation
function has been shown \cite{Collins2}. The cross sections for transversely
polarized photons was shown to be suppressed by $1/Q^2$ compared to the
longitudinal one. For the COMPASS kinematics in this analysis, the mean
$<Q^2>\approx 2$~GeV, which gives a ratio of $\sigma_L/\sigma_T\approx 1$. Further
studies will aim in a separation of the longitudinal and transverse part of the
cross section using the angular distribution of the $\rho^0 $ decay products
\cite{Diehl}. 

The current data taking at COMPASS with a transversely polarized $NH_3$ target
will provide the possibility to study  exclusive meson production on the
proton as well.


\begin{footnotesize}

\end{footnotesize}


\end{document}